\documentclass[aps,prl,twocolumn,superscriptaddress,amsmath]{revtex4-1}

\usepackage{graphicx}
\usepackage{dcolumn}
\usepackage{bm}
\usepackage{hyperref}
\usepackage{subfigure}
\usepackage{color}
\usepackage{siunitx}
\newcommand{\susi}[1]{\text{\tiny #1}}
\newcommand{\uni}[1]{\,\text{#1}}
\begin{document}

\title{Fast microwave beam splitters from superconducting resonators}

\author{M. Haeberlein}
\email[]{max.haeberlein@wmi.badw-muenchen.de}
\affiliation{Walther-Mei{\ss}ner-Institut, Bayrische Akademie der Wissenschaften, D-85748 Garching, Germany }
\affiliation{Physik-Department, Technische Universit\"{a}t M\"{u}nchen, D-85748 Garching, Germany}

\author{D. Zueco}
\affiliation{Instituto de Ciencia de Materiales de Arag\'on y Departamento de F\'isica de la Materia Condensada, CSIC-Universidad de Zaragoza, Pedro Cerbuna 12, 50009 Zaragoza, Spain}
\affiliation{Fundaci\'on ARAID, Paseo Mar\'ia Agust\'in 36, 50004 Zaragoza, Spain}

\author{P. Assum}
\affiliation{Walther-Mei{\ss}ner-Institut, Bayrische Akademie der Wissenschaften, D-85748 Garching, Germany }

\author{T. Wei{\ss}l}
\affiliation{Institut N\'eel, CNRS, F-38042 Grenoble cedex 9}

\author{E. Hoffmann}
\affiliation{Walther-Mei{\ss}ner-Institut, Bayrische Akademie der Wissenschaften, D-85748 Garching, Germany }
\affiliation{Physik-Department, Technische Universit\"{a}t M\"{u}nchen, D-85748 Garching, Germany}

\author{B. Peropadre}
\affiliation{Instituto de F{\'{\i}}sica Fundamental, IFF-CSIC, Serrano 113-B, E-28006 Madrid, Spain}

\author{J.J. Garc\'ia-Ripoll}
\affiliation{Instituto de F{\'{\i}}sica Fundamental, IFF-CSIC, Serrano 113-B, E-28006 Madrid, Spain}

\author{E. Solano}
\affiliation{Departamento de Qu\'imica F\'isica, Universidad del Pa\'is Vasco UPV/EHU, Apartado 644, 48080 Bilbao, Spain}
\affiliation{IKERBASQUE, Basque Foundation for Science, Alameda Urquijo 36, 48011 Bilbao, Spain}

\author{F. Deppe}
\affiliation{Walther-Mei{\ss}ner-Institut, Bayrische Akademie der Wissenschaften, D-85748 Garching, Germany }
\affiliation{Physik-Department, Technische Universit\"{a}t M\"{u}nchen, D-85748 Garching, Germany}

\author{A. Marx}
\affiliation{Walther-Mei{\ss}ner-Institut, Bayrische Akademie der Wissenschaften, D-85748 Garching, Germany }

\author{R. Gross}
\affiliation{Walther-Mei{\ss}ner-Institut, Bayrische Akademie der Wissenschaften, D-85748 Garching, Germany }
\affiliation{Physik-Department, Technische Universit\"{a}t M\"{u}nchen, D-85748 Garching, Germany}

\date{\today}

\begin{abstract}
Coupled superconducting transmission line resonators have applications in quantum information processing and fundamental quantum mechanics. A particular example is the realization of fast beam splitters, which however is hampered by two-mode squeezer terms. Here, we experimentally study superconducting microstrip resonators which are coupled over one third of their length. By varying the position of this coupling region we can tune the strength of the two-mode squeezer coupling from \SI{2.4}{\percent} to \SI{12.9}{\percent} of the resonance frequency of \SI{5.44}{GHz}. Nevertheless, the beam splitter coupling rate for maximally suppressed two-mode squeezing is \SI{810}{\mega\hertz}, enabling the construction of a fast and pure beam splitter.
\end{abstract}

\pacs{}

\maketitle

Recent advances in quantum electrodynamics with superconducting circuits (circuit QED) allowed for the experimental implementation of basic quantum computation algorithms~\cite{Lucero_prime_2012}.
 Based on important results such as single photon generation~\cite{hofheinz_generation_2008} and multi-qubit gates~\cite{dicarlo_demonstration_2009,fedorov_implementation_2012}, advanced schemes for quantum error correction~\cite{divincenzo_fault-tolerant_1996} and quantum feedback control~\cite{vijay_quantum_2012} were proposed.
 In such digital approaches, superconducting quantum circuits substitute classical bits and bus systems, allowing one to construct a general purpose quantum computation device.
 However, digital quantum simulations typically require a large number of qubits and sophisticated error correction schemes~\cite{buluta_quantum_2009}, which is still a significant technological challenge to date.
 Therefore, in the short term it is more promising to focus on what is called analog quantum computation or simulation.
 In this approach, a model quantum system  is used to set up a quantum mechanical evolution similar to the physical system of interest.
 However, contrary to the physical system, the input and output channels of the model system are easily accessible.
 Superconducting quantum circuits interacting with quantum microwave fields represent a particularly attractive model system~\cite{schoelkopf_wiring_2008}.
 If the microwave fields  are confined inside cavities, proposals and early experiments towards the simulation of manybody Hamiltonians exist~\cite{leib_bosehubbard_2010,houck_2012}.
 Beyond that, recent work on systems involving propagating quantum microwaves~\cite{menzel_path_2012,hoi_giant_2012} suggests that it is possible implement all-optical quantum simulation schemes~\cite{obrien_optical_2007} in the microwave regime.
 This route seems particularly attractive, because superconducting circuits offer extraordinarily large nonlinearities~\cite{niemczyk_circuit_2010} and therefore promise deterministic gates.
 A qubit can, for example, be encoded in an entangled state of two spatially separated superconducting waveguides.
 In such a situation, linear microwave beam splitters play an important role for the realization of single qubit rotations and two qubit Knill-Laflamme-Milburn gates~\cite{knill_scheme_2001,divincenzo_2010}.
 At this point, it is important to consider decoherence effects.
 In order to minimize them, a beam splitter should be fast in the sense that its coupling rate is a significant fraction of the frequency of the propagating microwaves.
 In such an ultrastrong coupling scenario, it is well-known~\cite{niemczyk_circuit_2010,kike_deep_2010} that nonlinear effects arise for dipolar coupling.
 Hence, these nonlinear effects must be taken care of in order to ensure a pure beam splitter functionality.
 In this work, we first develop a theoretical model for fast and pure microwave beam splitters based on two frequency-degenerate coupled superconducting transmission line resonators with low external quality factors.
 We confirm this model with proof-of-principle experiments using microstrip resonators with a resonance frequency of $\omega_\susi{0}{/}2\pi\,{=}\,$\SI{5.44}{\giga\hertz} and medium quality factors ranging between $150$ and $600$.
 Notably, we reach a beam splitter coupling strength of above \SI{800}{\mega\hertz} while suppressing the nonlinear coupling by a factor of six by exploiting the \SI{90}{\degree} phase shift between the inductive and the capacitive coupling channel.
 This allows for many operations within  decoherence times of superconducting tramsission line circuits~\cite{barends_mini_2011}.
We first introduce our model, which is based on Ref.~\onlinecite{peropadre_tunable_2012}.
As we aim at the realization of a pure beam splitter, the Hamiltonian describing our experimental system ideally should read as
\begin{equation}
\label{BShamiltonian}
{\cal H}= \hbar\omega_\susi{0} \left ( a^{\dagger}a + b^{\dagger}b \right ) +\hbar g_\susi{BS} \left ( a^{\dagger}  b + a b^{\dagger} \right).
\end{equation}
Here, $a^\dag$, $b^\dag$, $a$, and $b$ are the bosonic creation and annihilation operators of the two resonators and $g_\susi{BS}$ is the beam splitter coupling rate.
 The beam splitter interaction term $g_\susi{BS}(a^{\dagger}b\,{+}\,ab^{\dagger})$ splits the single resonance symmetrically, resulting in two new normal modes at the angular frequencies $\omega_{\pm}\,{=}\,\omega_\susi{0}\,{\pm}\,g_\susi{BS}$.
 We can apply Eq.~(\ref{BShamiltonian}) to the case of two transmission line resonators coupled in a small region around a position where either the magnetic field (current) or the electric field (voltage) has an antinode.
 While this scenario allows one to neglect either the capacitive or the inductive coupling channel, it limits practical devices to coupling rates smaller than approximately $g\,{/}\,\omega_\susi{0}\,{\approx}\,2\%$.
 In order to achieve higher coupling rates, we distribute the coupling over a region spanning a significant fraction of the resonator length.
 As a consequence of the large coupling strength, the rotating wave approximation breaks down, giving rise to a two-mode squeezer (TMS) term in the Hamiltonian. Introducing the TMS coupling rate $g_\susi{TMS}$, we find
 \begin{equation}
\label{fullhamiltonian}
{\cal H} = \hbar\widetilde{\omega} \left ( a^{\dagger}a + b^{\dagger}b \right ) + \hbar g_\susi{BS} \left( a^{\dagger} b + a b^{\dagger} \right)  + \hbar g_\susi{TMS} \left( a^{\dagger} b^{\dagger} + a b \right).
\end{equation}
 This Hamiltonian describes two coupled harmonic oscillators of renormalized frequency $\widetilde{\omega}$, which is split -- in general asymmetrically with respect to $\omega_\susi{0}$ -- into two normal modes of frequencies $\omega_\pm$. The detailed definition of $\widetilde{\omega}$ is not relevant for this work and can be found in
 Ref.~\onlinecite{peropadre_tunable_2012}.
 The total coupling rate results from a superposition of a capacitive ($g_{\rm c}$) and an inductive ($g_{\rm i}$) coupling channel.
 The corresponding two channels couple via $90^\circ$-shifted single mode fields.
 Therefore, we find $g_\susi{BS}\,{\equiv}\,\left( g_\susi{i}+g_\susi{c}\right)$ and $g_\susi{TMS}\,{\equiv}\,\left( g_\susi{i}-g_\susi{c}\right)$.
 The coupling rates $g_{\rm c}$ and $g_{\rm i}$ depend solely on the resonator geometry.
 For a suitable resonator design, we can achieve $g_{\rm c}=g_{\rm i}$ and thus $g_\susi{TMS}=0$.
 In other words, our distributed coupling approach allows for the realization of a pure beam splitter described by the Hamiltonian of Eq.~(\ref{BShamiltonian}) with large coupling rates $g_\susi{BS}$.
\begin{figure}
\includegraphics{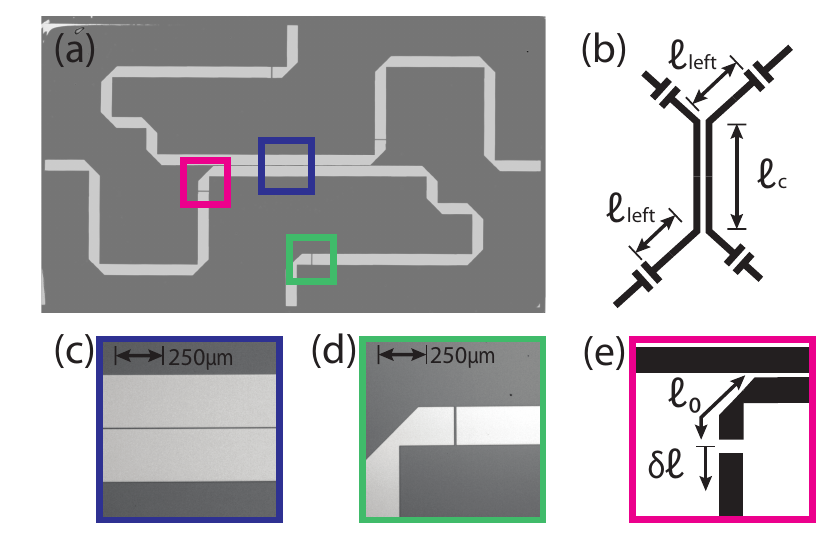}
\caption{\label{setup}
Sample layout. (a) Reworked photograph of two coupled resonators on a \SI{10x6}{\milli\metre} silicon wafer. (b) Schematic circuit diagram. The resonators couple over an electrical length $\ell_\susi{c}$. The coupling region starts at the electrical length $\ell_\susi{left}$.  (c, d) Enlarged view of the region marked with the blue (green) box in (a). (e) Sketch of the region marked with the purple box in (a).
}
\end{figure}

\begin{figure}
\includegraphics{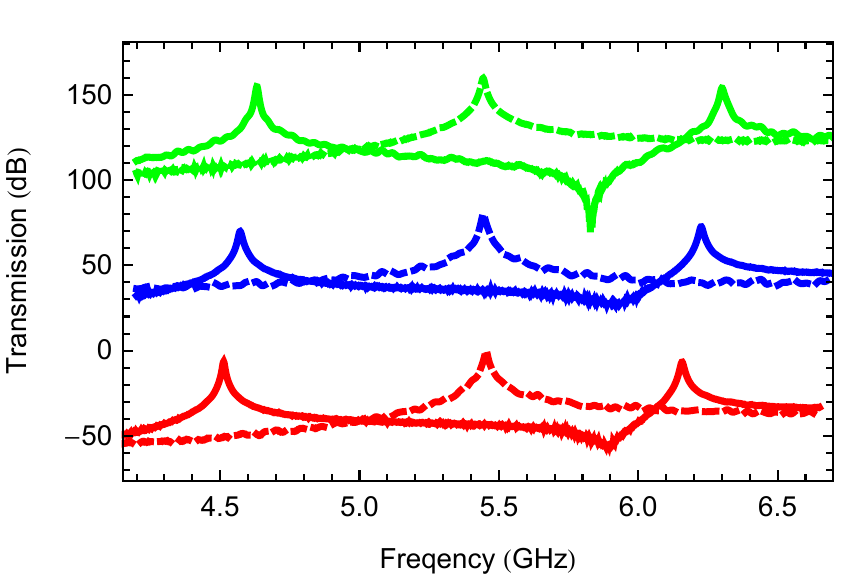}
\caption{\label{vnameas}
Transmission spectra of three resonators coupled over a length of \SI{3}{\milli\metre}, with $\delta\ell\,{=}\,$\SI{0}{\micro\metre} (bottom),  \SI{300}{\micro\metre} (middle), and \SI{600}{\micro\metre} (top). The top and middle curves are shifted by \SI{160}{\decibel} and  \SI{80}{\decibel}, respectively. The dashed lines show the respective single resonator transmission spectra.
}
\end{figure}

In the remainder of this work, we experimentally validate the distributed coupling model of Eq.~(\ref{fullhamiltonian}) by varying $g_{\rm c}$ and $g_{\rm i}$ in a controlled way.
 To this end, we fabricate samples containing two coupled microstrip resonators.
 Our design is shown in Fig.~\ref{setup}.
 For the fabrication of the chip shown in Fig.~\ref{setup}(a), we first sputter \SI{100}{\nano\meter} Niobium on both sides of a \SI{250}{\micro\meter} thick SiO$_2$ (\SI{50}{\nano\meter}) coated silicon wafer.
 One side is then patterned by optical lithography and reactive ion etching, the other one serves as ground plane.
 Our microstrip waveguides have a width of \SI{204}{\micro\meter} to match an impedance of \SI{50}{\ohm}.
 As shown in Fig.~\ref{setup}(a), the two resonators have the same shape.
 In order to avoid geometry effects, we shift the position of the  coupling capacitors defining both ends of the resonators rather than redesigning the coupling region.
 In this way, we investigate seven different configurations where the resonators are coupled over a length of $\ell_\susi{c}\,{=}\,3\uni{mm}$ at different physical coupling positions $\ell_\susi{left}$ [see Fig.~\ref{setup}(b)].
 For each two-resonator sample, we fabricate a single resonator sample with the same parameters for comparison.
 In our experiments, we measure transmission spectra of the fundamental mode of single and coupled microstrip resonators at $4.2\uni{K}$ and extract the resonance frequencies $\omega_\susi{0}$ and $\omega_{\pm}$, respectively.
 Typical examples are shown in Fig.~\ref{vnameas}.
 It can be seen that the two transmission peaks of the coupled resonators split asymmetrically with respect to the peak of the single resonator.
 This already indicates that the coupling is described by Eq.~\ref{fullhamiltonian}.
 Furthermore, we find that $\omega_\susi{0}\,{=}\,$\SI{5.44}{\giga\hertz} is independent of the position of the coupling capacitors as expected.
 Hence, also the total electrical length $\ell_\susi{tot}\,{=}\,\pi\/c\,{/}\,(\omega_\susi{0}\sqrt{\epsilon_\susi{eff}})$ is the same for all configurations.
 With $c\,{=}\,$\SI{2.99e8}{\meter/\second} and the effective dielectric constant $\epsilon_{\rm eff}\,{=}\,7.59$, we find $\ell_{\rm tot}\,{=}\,$\SI{9.963}{\milli\meter}.

\begin{figure}
\includegraphics{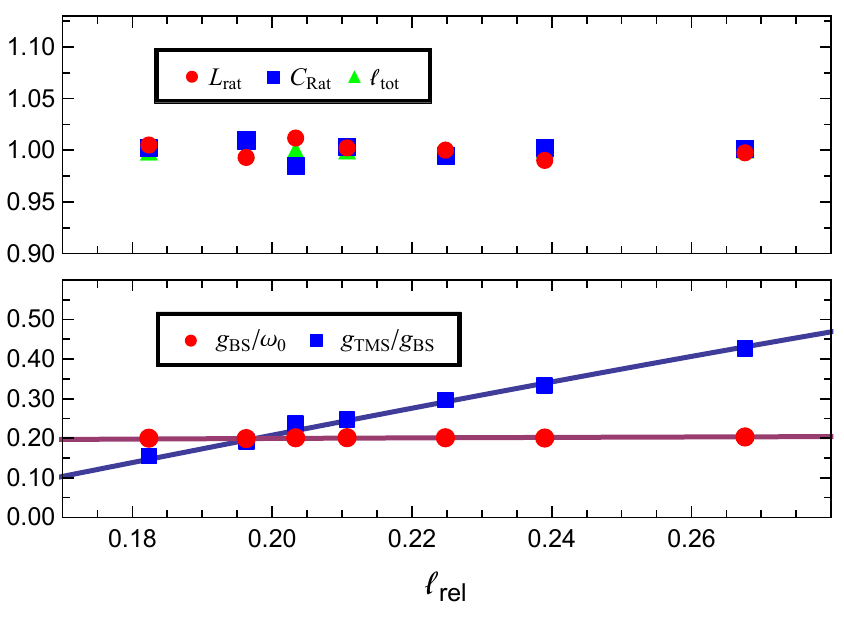}
\caption{\label{fitsnerrors}
Top: The parameters $L_\susi{rat}$, $C_\susi{rat}$, and $\ell_\susi{tot}$ divided by their respective average value displayed against the relative coupling position.
Bottom: Coupling ratios as a function of the relative coupling position. The solid line is obtained using the average values of $L_\susi{rat}$, $C_\susi{rat}$, and $\ell_\susi{tot}$.
}
\end{figure}

In order to extract the coupling parameters $g_\susi{c}$ and $g_\susi{i}$, we make use of the microscopic model~\cite{peropadre_tunable_2012} leading to Eq.~(\ref{fullhamiltonian}).
 The input parameters to this model are the ratio of the self-inductance (capacitance) per length to the mutual inductance (capacitance) per unit length $L_\susi{rat}$ ($C_\susi{rat}$) and the electrical position of the coupling region.
 The latter cannot be determined directly from the sample geometry because the physical length of the transmission line differs from the electrical one whenever there is a bend in the resonator.
 Furthermore, the ratios $L_\susi{rat}$ and $C_\susi{rat}$ depend implicitly on the electrical coupling position.
 Hence, the first step in our analysis is the determination of the electrical position of the coupling
 \begin{equation}
    \label{lleft}
    \ell_\susi{left} = \ell_\susi{0} + \delta\ell.
 \end{equation}
 Here, as shown in Fig.~\ref{setup}(e), $\ell_\susi{0}$ is the minimum distance between the coupling capacitor and the border of the coupling region and $\delta\ell$ accounts for the varying position of the coupling capacitor.
 We obtain $\delta\ell$ directly from the resonator geometry because in good approximation the electrical length of a straight segment of the resonator equals its physical length.
 With the definition of Eq.~(\ref{lleft}) and the model in Ref.~\onlinecite{peropadre_tunable_2012}, we can write
 \begin{eqnarray*}
    & & \\
    L_\susi{rat} & = & \frac{\nu_\susi{L} \omega_\susi{+}^2}{\omega_\susi{0}^2-\nu_\susi{L} \omega_\susi{+}^2} \\
    C_\susi{rat} & = &  \frac{\nu_\susi{C}(\omega_\susi{0}^2-\omega_\susi{-}^2) + \nu_\susi{C} \omega_\susi{0}^2 \omega_\susi{+}^2/(\omega_\susi{0}^2-2 \nu_\susi{L} \omega_\susi{+}^2)}{2\omega_\susi{-}^2}.
 \end{eqnarray*}
 Here, $\nu_\susi{L,C}\,{=}\,\ell_\susi{tot}/\Delta_\susi{L,C}$ are geometry factors.
 The expressions $\Delta_\susi{L,C}$ represent the overlap integrals of the magnetic (electric) field modes.
 For our scenario of homogeneous resonators and fundamental mode coupling, we can set $2\pi\Delta_\susi{L,C}=\ell_\susi{c}\,{\mp}\, \ell_\susi{tot}\left[\textrm{sin}(2\pi(\ell_\susi{left}{+}\ell_\susi{c})/\ell_\susi{tot}) - \textrm{sin}(2\pi\ell_\susi{left}/\ell_\susi{tot})\right]$.
 In order to extract $\ell_\susi{0}$ from the measured peak positions $\omega_\susi{0}$ and $\omega_\pm$, we first assume that the field in the resonators is a TEM-mode and, consequently, $L_\susi{rat}$ and $C_\susi{rat}$ are independent of $\ell_\susi{left}$.
 Subsequent minimization of the normalized variance of $L_\susi{rat} + C_\susi{rat}$ for all seven capacitor configurations yields $\ell_\susi{0}\,{=}\,$\SI{1.271}{\milli\meter}.
 Figure \ref{fitsnerrors}(a) shows that indeed for this value of $\ell_\susi{0}$, the parameters $L_\susi{rat}$ and $C_\susi{rat}$ do not deviate more than 
 \SI{3}{\percent} from their average value.
 This gives evidence that our model is self-consistent.

In the next step, we use $L_\susi{rat}$, $C_\susi{rat}$, and $\ell_\susi{0}\,{+}\,\delta\ell$ to calculate $g_\susi{BS}$, $g_\susi{TMS}$, and $\tilde{\omega}$.
 In Fig.~\ref{fitsnerrors}(b), we show $g_\susi{BS}/\omega_\susi{0}$ and $g_\susi{BS}/g_\susi{TMS}$ as a function of $\ell_\susi{rel}\,{\equiv}\,\ell_\susi{left}/(\ell_\susi{tot}\,{-}\,\ell_\susi{c})$.
 We observe a maximum suppression of the TMS coupling rate to $g_\susi{TMS}/g_\susi{BS}\,{=}\,16\%$ and a minimum suppression of $g_\susi{TMS}/g_\susi{BS}\,{=}\,43\%$ while the beam splitter coupling rate stays nearly constant at  $g_\susi{BS}\,{=}\,$\SI[separate-uncertainty]{816+-7}{\mega\hertz}.
 An extrapolation of the model prediction suggests that the TMS coupling should vanish at the relative coupling position $\ell_\susi{rel}\,{=}\,14\uni{\%}$.
 Nevertheless, the beam splitter coupling rate at this position still exceeds \SI{780}{\mega\hertz}.
 This configuration is ideally suited for the realization of a fast beam splitter and can in principle be reached with our geometry.
Finally, we analyze the potential of our devices for the investigation of ultrastrong coupling.
 In this context, we note that the coupling between the two resonators can be ultrastrong in the same way as the qubit-resonator coupling discussed in Ref.~\onlinecite{niemczyk_circuit_2010}.
 For our samples, we achieve a maximum TMS coupling rate of \SI{351}{\mega\hertz} for $\ell_\susi{rel}\,{=}\,27\%$.
 When moving the coupling region to the center of the resonators, the maximum rate would become \SI{702}{\mega\hertz} and $g_\susi{TMS}/\omega_\susi{0}\,{=}\,12.9\%$.
 This implies that the relative coupling rate of our device is equally large as in Ref.~\onlinecite{niemczyk_circuit_2010}.
 Since our devices do not require nonlinearities and are therefore much easier to fabricate, they provide a promising way of studying the dynamics of ultrastrong coupling.

In summary, we use linear superconducting circuits to implement a Hamiltonian with a beam splitter coupling strength of more than \SI{800}{\mega\hertz}, where the TMS term is suppressed by a factor of six.
 We demonstrate a tunability of the coupling ratio $g_\susi{TMS}/g_\susi{BS}$ between \SI{16}{\percent} and \SI{43}{\percent}.
 An extrapolation of our result shows that an ultrastrong coupling scenario as well as a pure beam splitter Hamiltonian can be reached with our sample design.
 This paves the way for studying ultrastrong coupling dynamics and, by design of a suitable capacitor configuration~\cite{divincenzo_2010}, building fast beam splitter circuits for analog quantum computation and simulation with both standing-wave and propagating quantum microwaves.

\begin{acknowledgments}
We acknowledge support from the Deutsche Forschungsgemeinschaft via SFB 631,
the German excellence initiative via the ‘Nanosystems Initiative Munich’ (NIM), from the EU
projects SOLID, CCQED and PROMISCE, EPSRC EP/H050434/1, Basque Government IT472-10, and
Spanish MINECO FIS2009-12773-C02-01, FIS2011-25167, and FIS2012-36673-C03-02.
\end{acknowledgments}

\bibliographystyle{apsrev4-1}

\end{document}